# Desorption of an active Brownian polymer from a homogeneous attractive surface

Guo-qiang Feng[a], Wen-de Tian[a]*

[a] *Center for Soft Condensed Matter Physics & Interdisciplinary Research, School of Physical Science and Technology, Soochow University, Suzhou 215006, China*

*Corresponding authors: tianwende@suda.edu.cn (W.-d. Tian)

**Abstract**

The interfacial behavior of a flexible polymer with activity, named active Brownian polymer (ABPO), is studied by Langevin dynamics simulations. On the dependence of adsorption strength and activity characterized by Péclet number (*Pe*), the polymer displays two typical states on the surface: adsorption and desorption states. We find the diffusion behavior of ABPO parallel to the surface yields the "active Rouse model" and activity causes the adsorption-desorption transition at a certain adsorption strength. Particular attention is paid to how the desorption time, $\tau_{des}$, changes with the activity. At intermediate activity, $\tau_{des}$ displays an exponential decay with the inverse of effective temperature, $T_{eff} \propto 1 + Pe^2/18$, reminiscent of the mechanism of thermal activation. At higher activity, due to easily overcoming the attractive energy barrier, $\tau_{des} \propto Pe^{-1}$ is found. At lower activity, a power-law dependence of $\tau_{des}$ on the diffusion coefficient perpendicular to the surface ($D_\perp$) is observed ($\tau_{des} \sim D_\perp^{-1.28}$). Further, we observed a non-monotonic dependence of desorption time on the rotation diffusion coefficient $D_R$ of monomer and found $\tau_{des}$ exists a scaling relation with chain length *N*, $\tau_{des} \sim N^\phi$, and the scaling exponent $\phi$ decreases with the increase of activity. Our results highlight the activity can be used to regulate the polymer adsorption and desorption behavior.

## 1. Introduction

Active matter, whose agent converts chemical energy or other forms of energy into mechanical motion, is inherently out of equilibrium.[1-12] Understanding behavior of active systems is an intense area of research. One of the widely studied systems is active polymers. Various types of active polymers can be found in the living system.[13, 14] For example, microtubules, propelled via self-tread-milling or molecular motors in the ATP solution, are ubiquitous in the cell.[15] Actin filaments can be driven by motor proteins anchored on a substrate.[16] DNA is tuned by bio-enzymes such as DNA polymerase and helicases.[17] Besides, active "colloidal polymers" have been synthetized in laboratory to study the non-equilibrium dynamics of chains.[18-20] Currently, two kinds of active polymer models have been developed.[13] One is the self-propelled filament motivated by the active bio-filaments in motility assays.[21-23] The other is the active Brownian polymer (ABPO), which captures the main feature of "polymers" composed of active agents or passive polymers embedded in a bath of active fluid.[24-26]

The ABPO consisting of active Brownian particles (ABPs) exhibits numerous interesting features that are usually absent in traditional polymers. For example, ABPO can shrink and then swell with the increase of activity in three-spatial dimensions.[24] The behavior is attributed to the entropic penalty to chain stretching, which results from energy compensation by activity. Although the structure and dynamics of ABPOs in the two and three dimensions drew immense interest,[27-30] less attention was paid to their interfacial behaviors. The motivation is threefold: first, polymer behavior on the surface, closely related to technological applications such as polymer solubilizers and colloidal stabilization, is fundamental in polymer physics.[31-33] Second, active biopolymers are ubiquitous in biological systems, in which the interfaces are rich because of the existence of massive organelles. Third, from a fundamental point of view, it is important to know the active effect on the polymer adsorption and desorption, which might be different from that in the equilibrium state. Understanding the interfacial behavior of non-equilibrium polymer is helpful in the design of self-healing biomaterials.

Adsorption/desorption is a ubiquitous phenomenon affecting the properties of interfaces. For the traditional polymer, experimental and theoretical studies[34-37] have been devoted to exploring the desorption affected by adsorption strength,[38-40] surface roughness,[41-43] and crowding environment.[44, 45] For instance, Luo et al.[46] found the interplay of external driving force and thermal noise results in an adsorption-desorption transition. Paturej et al.[47, 48] obtained a scaling relation of $\langle \tau_d \rangle \propto N^2$ between detachment time, $\langle \tau_d \rangle$, and polymer length, $N$, for the force-assisted desorption. Without external force, the desorption is driven by an increasing configurational entropy of chain, which is against the enthalpy originating from polymer-surface attraction. In contrast to an external force, the activity on ABPO is along the inherent orientation of each monomer. For biological system, the activity might originate from several physical and chemical driving forces such as active coupling due to electrostatic interactions and ATP hydrolysis.[49, 50] Hence, several interesting questions are worth considering: 1) What is the structure and dynamics of ABPO on a surface? 2) Could the polymer be detached from an attractive surface when turning on active force? 3) Are there any differences of desorption dynamics with passive polymers?

To address the above questions, a series of Langevin dynamics (LD) simulations were performed to investigate the conformational and dynamical behaviors of ABPO at a flat attractive surface. Special attention was paid to the effect of the active force, attractive strength, and chain length. The adsorption state appears at large attractive strength and low active force, the desorption state arises at small attractive strength and high active force. At a certain adsorption strength, the increase of active force leads to the adsorption-desorption transition. Further, we observed an exponential dependence of desorption time $\tau_{des}$ on the inverse of effective temperature, $1/T_{eff}$, at intermediate activities which can be understood by effective thermally-activated process. What's more interesting, the deviation of $\tau_{des}$ from the exponential law at low and high activity was also found. Finally, $\tau_{des}$ increases in a power-law relation with chain length, $N$, with the scaling exponent decreasing with activity and desorption time shows a non-monotonic dependence on the rotational diffusion coefficient.

This paper is organized as follows. In Section 2, the model and simulation details are introduced. Results and discussion are given in Section 3. Finally, the conclusion is summarized.

## 2. Model and methods

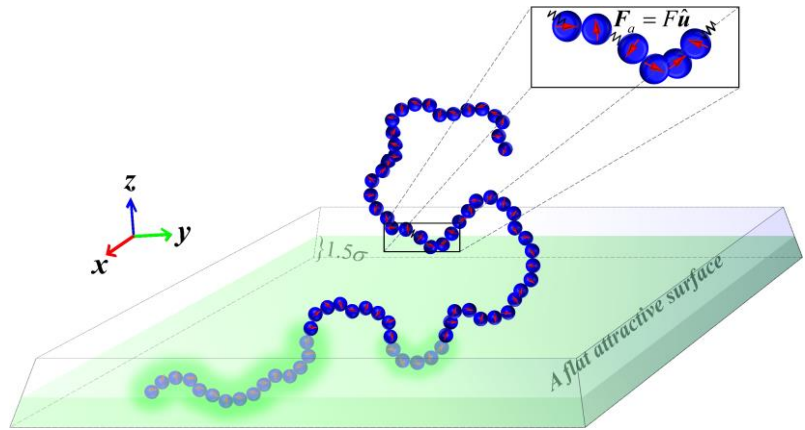

Fig. 1. Schematic diagram of an active Brownian polymer (ABPO) on a flat attractive surface (green). Red arrows represent the orientation of active force on monomers (blue beads). A monomer above the surface with distance larger than $1.5\sigma$ is considered to be detached.

We performed LD simulations of a flexible ABPO consisting of $N$ ABPs on a flat attractive surface. We used a standard bead-and-spring model, which includes elastic springs connecting consecutive beads and a Weeks-Chandler-Anderson (WCA) potential to mimic steric interactions among monomers. Similar to our previous studies,[19, 51-53] each bead has an intrinsic orientation, $\hat{u}$ (see Fig.1). The active force exerted on each bead is $\boldsymbol{F}_a = F\hat{\boldsymbol{u}}$, where $F$ is a constant magnitude. The ABPO was placed above the surface (at $z=0$), which is modeled by a flat attractive smooth plane.

Total interaction, $U$, includes non-bonded energy, $U_{nb}$, bonding energy, $U_b$, of adjacent bonded beads, and bead-surface energy, $U_s$.

$$U = U_{nb} + U_b + U_s \tag{1}$$

Here, WCA potential is used to describe the non-bonded interaction of all beads, which is given by

$$U_{nb} = \sum_{i<j}^{N} U(r_{ij}), \quad U(r_{ij}) = \begin{cases} 4\varepsilon\left[\left(\dfrac{\sigma}{r_{ij}}\right)^{12} - \left(\dfrac{\sigma}{r_{ij}}\right)^{6}\right] + \varepsilon & r_{ij} \leq 2^{1/6}\sigma \\ 0 & r_{ij} > 2^{1/6}\sigma \end{cases} \quad (2)$$

where $r_{ij}$, $\varepsilon$, $\sigma$ represent the center-to-center distance between the $i$ th and $j$ th bead, potential strength, and diameter of bead, respectively. The bonding energy is described by

$$U_b = \frac{1}{2} k \sum_{i=1}^{N-1} (r_i - r_0)^2 \quad (3)$$

where $k = 2000.0 k_B T / \sigma^2$ is elastic coefficient, $r_i$ the $i$ th bond length, and $r_0 = \sigma$ the equilibrium bond length.

The bead-surface interaction is given by a truncated 9-3 Lennard-Jones (LJ) potential,

$$U_s = \sum_{i=1}^{N} U(z_i), \quad U(z_i) = \begin{cases} \varepsilon_s \left[\dfrac{2}{15}\left(\dfrac{\sigma}{z_i}\right)^9 - \left(\dfrac{\sigma}{z_i}\right)^3\right] & z_i \leq z_{lower} = 3.0\sigma \\ 0 & \text{otherwise} \end{cases} \quad (4)$$

where $\varepsilon_s$ is the adsorption strength as an adjustable parameter in our simulations and $z_i$ is the distance between the $i$ th bead and the surface.

The dynamics of ABPO is described by Langevin equation below.

$$m\ddot{\mathbf{r}} = -\gamma_T \dot{\mathbf{r}} - \nabla U + \mathbf{\Gamma} + \mathbf{F}_a, \quad \dot{\hat{\mathbf{u}}} = \boldsymbol{\zeta} \times \hat{\mathbf{u}} \quad (5)$$

Where $m$ is monomer mass, $\gamma_T$ the translational friction coefficient, $\mathbf{\Gamma}$ the translational thermal noise and $\boldsymbol{\zeta}$ the rotational noise. They obey the fluctuation-dissipation theorems.

$$\langle \mathbf{\Gamma}(t) \cdot \mathbf{\Gamma}(t') \rangle = 2d\gamma_T^2 D_T \delta(t-t'), \quad \langle \boldsymbol{\zeta}(t) \cdot \boldsymbol{\zeta}(t') \rangle = 2(d-1)D_R \delta(t-t') \quad (6)$$

Here $d$ is the dimension of system. $D_T = k_B T / \gamma_T$ is the translational diffusion coefficient, $k_B$ the Boltzmann constant, $T$ the temperature. $D_R$ is the rotational diffusion coefficient, and we fixed $D_R = 3D_T / \sigma^2$ unless noted otherwise.

We used the home-modified LAMMPS molecular dynamics software[54] to perform all simulations. The equations of motion were solved using a velocity-Verlet algorithm. Periodic boundary conditions (PBCs) were applied in the $x$ and $y$ directions and the fixed boundary condition was applied in the $z$ direction. The box size is $N\sigma \times N\sigma \times 2N\sigma$. To avoid the ABPO escaping from the $z$ direction, there is an impenetrable wall at

$z = 2N\sigma$. Reduced units were used by setting $m = 1.0$, $\sigma = 1.0$ and $\varepsilon = k_B T = 1.0$. The reduced time unit is $\tau = \sqrt{k_B T/(m\sigma^2)}$. We fixed $\delta t = 2.0 \times 10^{-3} \tau$, $\gamma_T = 1.0 \times 10^2$ (to guarantee that the system is overdamped).[55-59] We mainly focused on the effects of active force, attractive strength, and chain length. The strength of active force can be characterized by a dimensionless Péclet number ($Pe$),[21] which is defined as $Pe = F\sigma/k_B T$. All simulations were carried out with the same procedures. Before turning on active force, ABPO was relaxed on the attractive surface with a long run of $2.0 \times 10^4 \tau$. Then a long-time ($4.0 \times 10^4 \tau$) simulation was performed with various active forces. The trajectory was saved every $2\tau$ for data analysis. To reduce the statistical error, independent simulations with different random numbers and initial configurations were up to $1.0 \times 10^3$ for each parameter. For comparison, we also studied the ABPO in two- and three-dimension boxes with PBCs in all directions.

3. Results

To analyze the dynamics of desorption processes, we defined several time-dependent quantities of ABPO such as the components ($R_{g\parallel}^2(t)$ and $R_{g\perp}^2(t)$) of mean-square radius of gyration, the components ($MSD_\parallel(t)$ and $MSD_\perp(t)$) of mean square displacement (MSD) of its center-of-mass, the number of detached monomer ($\mathcal{M}(t)$) and the correlation function ($G(t)$). They were calculated using the following equations:

$$\begin{aligned}
x_{cm}(t) &= \frac{1}{N}\sum_{i=1}^{N} x_i(t),\ y_{cm}(t) = \frac{1}{N}\sum_{i=1}^{N} y_i(t),\ z_{cm}(t) = \frac{1}{N}\sum_{i=1}^{N} z_i(t) \\
R_{g\parallel}^2(t) &= \left\langle \frac{1}{N}\sum_{i=1}^{N}\left\{\left[x_i(t)-x_{cm}(t)\right]^2 + \left[y_i(t)-y_{cm}(t)\right]^2\right\}\right\rangle \\
R_{g\perp}^2(t) &= \left\langle \frac{1}{N}\sum_{i=1}^{N}\left[z_i(t)-z_{cm}(t)\right]^2\right\rangle \\
MSD_\parallel(t) &= \left\langle \left(x_{cm}(t+t_0)-x_{cm}(t_0)\right)^2 + \left(y_{cm}(t+t_0)-y_{cm}(t_0)\right)^2\right\rangle \\
MSD_\perp(t) &= \left\langle \left(z_{cm}(t+t_0)-z_{cm}(t_0)\right)^2\right\rangle \\
\mathcal{M}(t) &= \left\langle M(t)\right\rangle \\
G(t) &= \left\langle \frac{N-M(t)}{N-M(0)}\right\rangle
\end{aligned} \quad (7)$$

where $r_i(t) = [x_i(t), y_i(t), z_i(t)]$ and $r_{cm}(t) = [x_{cm}(t), y_{cm}(t), z_{cm}(t)]$ represent the coordinate of the $i$ th bead and the center-of-mass of ABPO at time $t$, respectively. $t_0 = 2.0 \times 10^4 \tau$. $\langle ... \rangle$ denotes the ensemble average over the $1.0 \times 10^3$ independent trajectories. A monomer far from the surface with distance larger than $1.5\sigma$ is considered to be detached. $M(0)$ is the number of detached monomers at time $t = 0$; $M(t)$ is the number of detached monomers at time $t$. It should be noted that, when calculating $\mathcal{M}(t)$ and $G(t)$, $M(t)$ was set to be $N$ after the first full desorption of ABPO, i.e., as long as all monomers are detached, the subsequent re-adsorption of monomers were not included in the analysis. The desorption time $\tau_{des}$ was estimated from the exponential decay of the correlation function $G(t)$ with $G(t) \sim \exp(-t/\tau_{des})$.[35, 60-62]

Besides, we also calculated the time-independent quantities such as $\langle R_{g\parallel}^2\rangle$, $\langle R_{g\perp}^2\rangle$, and the average number of detached monomers $\langle M\rangle$ to investigate the stationary structure of ABPO. To calculate these quantities, trajectories of the last $10^4\tau$ were used for further reducing statistical errors. Meanwhile, the re-adsorbed monomers were included in $\langle M \rangle$. The re-adsorption behavior (Fig.S1) and the difference (Fig.S2) between $\mathcal{M}(t)$ and $\langle M \rangle$ are elaborated in Supporting Information (SI).

Generally, Van der Waals interaction (characterized by adsorption strength, $\varepsilon_s$) between polymer and interface is the driven force for polymer adsorption. From a thermodynamic point of view, a high temperature, which enlarges the weight of entropic effects, could cause polymer desorption. For the ABPO, propelling force is another interesting factor that activates the desorption. Due to the mutual competition of thermal noise and attractive force, here we fixed the thermal noise and focused on the influence of polymer activity (*Pe*).

### 3.1 Effect of *Pe*

We first focus on the structure and dynamics of ABPO near the surface with $N=100$ and $\varepsilon_s=10$. $\langle M \rangle$ as a function of *Pe* is plotted in Fig.2. At small *Pe*, total monomers are adsorbed on the surface without any detachment (see Fig.2). With the increase of *Pe*, part of the polymer is detached without complete desorption. At large *Pe*, the chain is fully desorbed and moves far away from the surface. $\langle M \rangle$ also shows a sharp increase at the intermediate *Pe*s between 20 and 40. The fluctuation of $\langle M \rangle$ is large in this intermediate region (the inset of Fig.2). Thus, the adsorption-desorption transition takes place as *Pe* is increased.

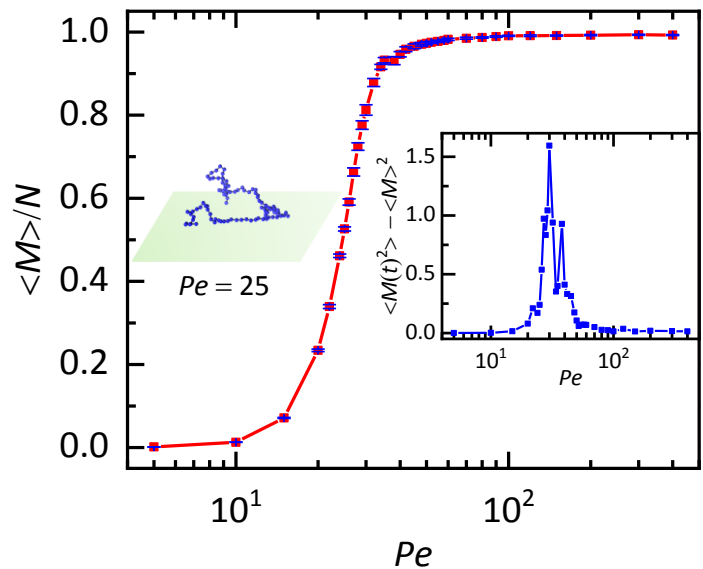

Fig. 2. The average desorbed number $\langle M \rangle$ as a function of *Pe* at $N=100$, $\varepsilon_s=10$. The insets display the deviation of desorbed number and typical snapshot of our system at *Pe* = 25.

The effect of active force on the structure of ABPO can be characterized by mean-square radius of gyration ($\langle R_g^2 \rangle$), corresponding to the size of a polymer. Its projection $\langle R_{g\parallel}^2 \rangle$ as a function of $Pe$ are calculated and plotted in Fig.3a. $\langle R_{g\parallel}^2 \rangle$s of the ABPO in two-dimensional (2D) and three-dimensional (3D) systems are also given. The desorption induced by active force can be manifested by the $\langle R_{g\parallel}^2 \rangle$ falling off from its 2D to 3D values with increasing active force in the region $20 < Pe < 30$, where an obvious shrinkage of the polymer appears in $xy$ dimensions (Fig.3a). At $Pe \leq 20$, the ABPO swells on the attractive surface, similar to the behaviors of ABPO in the 2D system. At $Pe > 30$, the swelling behavior of ABPO with the increase of $Pe$ is in line with that in the bulk. The collapse of ABPO in $xy$ dimensions at $20 < Pe < 30$ originates from the detachment of polymer monomers, which can also be manifested by the monotonous increase of $\langle R_{g\perp}^2 \rangle$ (as shown in Fig.S3 of SI). It should be noted that the steric interaction among monomers is important for the appearance of collapse of ABPO, which was not found for the active "phantom" polymers (data not shown).

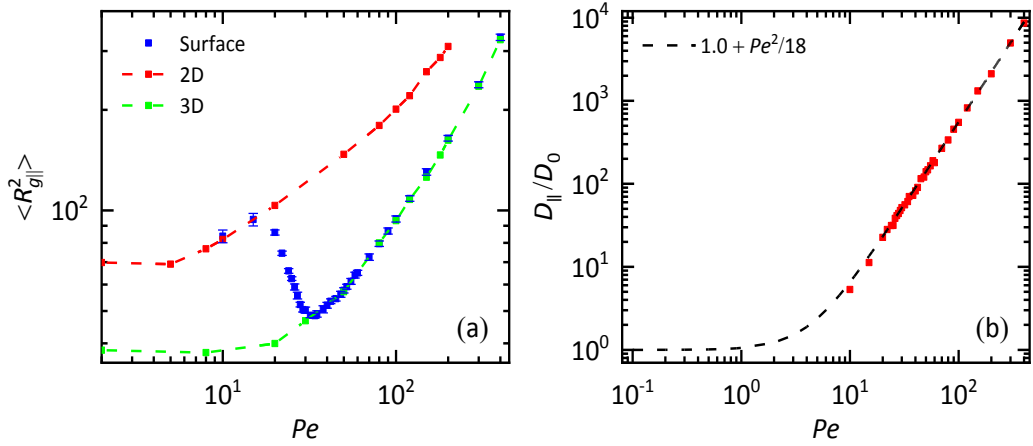

Fig. 3. $\langle R_{g\parallel}^2 \rangle$ (a) and the long-time diffusion coefficients $D_\parallel$ (b) as a function of $Pe$ and at $N = 100$, $\varepsilon_s = 10$. For comparison, the corresponding sizes in two-dimensional (2D) system and three-dimensional (3D) bulk are plotted in red and green squares, respectively. The $D_\parallel$ is normalized by $D_0 = k_B T / \gamma_T N$, which is the center-of-mass diffusion coefficient of passive polymer without active noise. The dashed line represents $D/D_0 = (1.0 + Pe^2/18)$ for 3D.

The diffusive behavior of ABPO was investigated via the $MSD$ of its center-of-mass and the diffusion coefficients $D_\parallel$ and $D_\perp$ were extracted by fitting $MSD_\parallel(t)$ and $MSD_\perp(t)$ at long-time scale according to Einstein relation $MSD(t) \sim 2dD_{eff}t$ (see Fig.3b and Fig.S4). The super-diffusive behavior at short time scale and normal diffusion at long time scale was observed in $xy$ directions for all $Pe$s, regardless of whether ABPO is detached or not (plotted by Fig.S4(a-b) and Fig.S5). The effective diffusion coefficient of ABPO $D_{eff}$ in the 3D system can be expressed as $D_{eff} = D_0(1 + Pe^2/18)$, in good agreement with "active Rouse model" [63-65] (as shown in Fig.3b). Thus, we define the effective temperature of the center of mass of the polymer $T_{eff} = T(1 + Pe^2/18)$, where $T$ is the background temperature from the thermal noise. It should be noticed that the effective temperature is different with the "kinetic temperature" ($\langle \dot{\boldsymbol{r}}^2 \rangle$) of each monomer[66], which is very lower relative to the $T_{eff}$ we defined. Due to the surface's attraction, the $D_\perp$ is smaller than $D_0(1+Pe^2/18)$ for $Pe<35$ (see Fig.S4c).

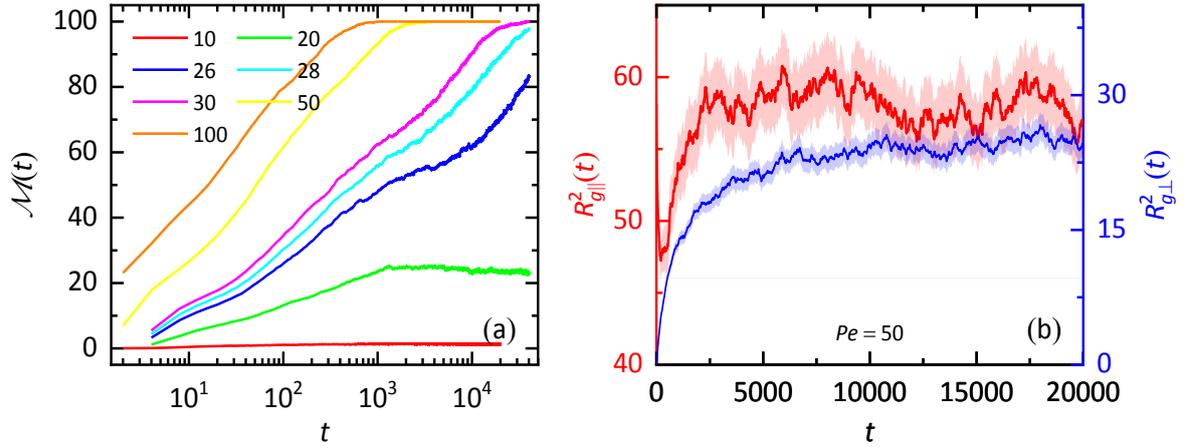

Fig. 4. (a) Time evolution of detached number $\mathcal{M}(t)$ for various $Pe$s and (b) $R_{g\parallel}^2(t)$ and $R_{g\perp}^2(t)$ as a function of simulation time at $Pe=50$, $N=100$, $\varepsilon_s=10$. The shaded error bar is also shown.

Next, we turn to the desorption dynamics of ABPO caused by the active force. The desorption process of ABPO is characterized by the time-evolution detached number $\mathcal{M}(t)$ as shown in Fig.4a. At $Pe=10$, the ABPO adhered to the attractive surface and almost no monomers are far from the surface in the time scale of our simulation. At the intermediate $Pe$, $\mathcal{M}(t)$s increase with time, but only partial monomers of ABPO depart from the surface, which means the chain does not desorb in the time window of our simulation. When the $Pe$ is beyond 30, more and more monomers are detached with the increase of simulation time and finally a complete desorption reaches in the time scale of our simulations. A typical time evolution of polymer size in $xy$ and $z$ direction for the system is given in Fig.4b. $R_{g\perp}^2(t)$ increases with time and finally approaches a plateau. Meanwhile, $R_{g\parallel}^2(t)$ decreases firstly due to partial monomers detaching from the surface, then increases with time due to further adjustment of polymer configuration aroused by active force, and finally also approaches a plateau, where the polymer is fully detached.

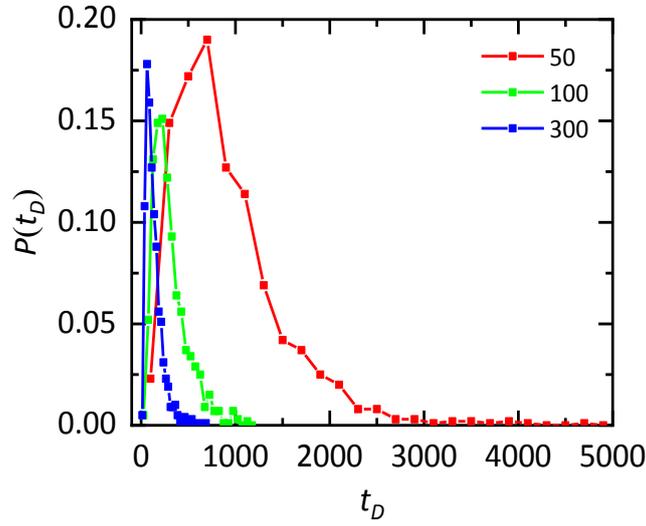

Fig. 5. Normalized probability distribution $P(t_D)$ of detachment time $t_D$ for $Pe = 50, 100, 300$ at $N = 100$, $\varepsilon_s = 10$.

Fig.5 shows the normalized probability $P(t_D)$ of $t_D$ required for the chain detaching from the surface at $Pe = 50, 100, 300$. For the three cases, the ABPO chain is complete desorption for all $1.0 \times 10^3$ independent simulations. The detachment time $t_D$ for each trajectory is determined by tracking $M(t)$. Once $M(t) = 100$, the time point is considered to be $t_D$. The probability displays a prominent peak followed by a tail, which indicates that only a small part of $1.0 \times 10^3$ independent simulations have a long detachment time. The larger $Pe$ leads to a short detachment time $t_D$. We also define the desorption time $\tau_{des}$, which is different with $t_D$, through fitting the correction time $G(t)$. The semi-log plot of $G(t)$ for various $Pe$s is given in Fig.6a. The behavior of $G(t)$ consists of a short transient period, where desorption is rapid, followed by an exponential decay as $G(t) \sim \exp(-t/\tau_{des})$ at long time scale. The transient period is related to the conformation transition after switching on the active force. The transient period decreases with the increase of active force, implying the desorption of monomers mainly controlled by conformation relaxation for large active force. The re-adsorption of ABPO on the surface is not considered in our $\mathcal{M}(t)$, the stretched-exponential decay[67] is not observed in our analysis. The advantage is that $G(t)$ only depends on desorption dynamics without re-adsorption and diffusion in the bulk. The linear region was fitted to obtain the desorption time $\tau_{des}$.

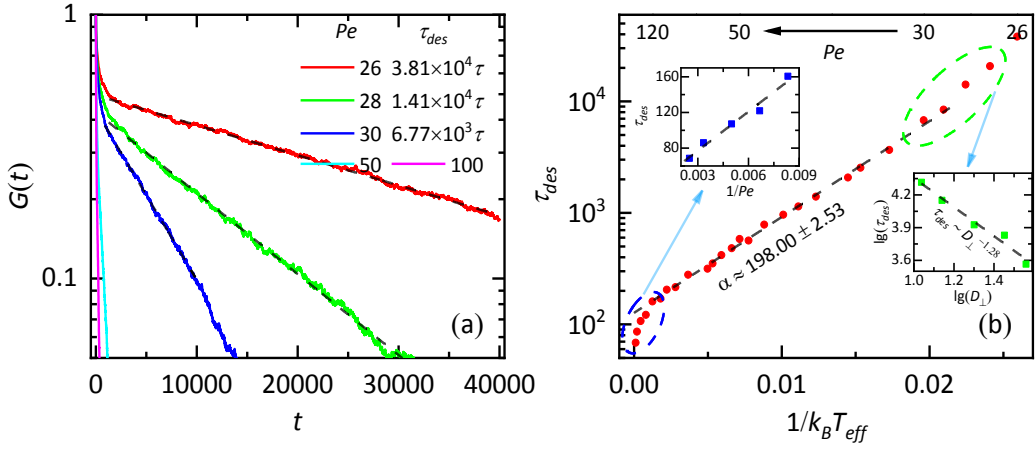

Fig. 6. (a) The decay of $G(t)$ for various $Pe$s at $N = 100$, $\varepsilon_s = 10$. The black dashed lines demonstrate the linear regime used to extract the desorption time $\tau_{des}$ via fitting $\lg G(t) \sim t$. (b) $\tau_{des}$ as a function of $T_{eff}$. The black dashed line represents the linear fitting of $\tau_{des} \sim \exp(\alpha/k_B T_{eff})$ with fitting parameter, $\alpha$, in the regime $32 \le Pe \le 120$. The green ($27 \le Pe \le 32$) and blue ($150 \le Pe \le 400$) circles denote regimes deviating from exponential decay.

Here, we only present desorption times for the cases of $Pe \ge 26$ in Fig.6b. To understand the desorption mechanism induced by active force, the dependence of $\tau_{des}$ on effective temperature $T_{eff}$ is semi-logarithmically plotted, and three regimes can be obtained. For $32 \le Pe \le 120$, the desorption time displays an exponential dependence on the $1/T_{eff}$, i.e., $\tau_{des} \sim \exp(\alpha/k_B T_{eff})$. Here $\alpha$ is a fitting parameter that depends on the polymer-surface interaction.[67] The process, similar to a particle (the center of mass of ABPO in our case) escaping over a potential barrier, often referred to as the Kramers' problem.[68] The active version, recently, has been studied by Caprini et al.[69] and Woillez et al.[70]. Caprini et al.[71], using the active Ornstein-Uhlenbeck model and the unified colored noise approximation, suggests that the average escape time of an active particle with small persistence exponentially decreases as $Pe$ is increased in near equilibrium regimes. The active force can be recast onto an effective potential with an effective diffusion coefficient, $D_{eff}$ ($\propto Pe^2$). In our model, the persistence time of ABPs is about $16.7\tau$, which is very shorter than the desorption time. For single ABP, the escape time obeys the analytical formula, $\ln(\tau_{des}) \sim 1/Pe^2$.[71] Though the existence of configuration entropy and correlated effect of ABPs linked by bonds, we still observed an exponential dependence of desorption time on the effective temperature at the intermediate $Pe$s. The desorption by thermal activation is also observed in passive polymer system in the limit of strong adsorption.[72] The difference is that the temperature they used is for the single monomer.

For larger $Pe$s ($\ge 150$), the desorption time decreases sharply with increasing $Pe$, in contrast to the exponential decay at intermediate $Pe$s. It can nearly be fitted by $\tau_{des} \propto Pe^{-1}$ (see the inset of Fig.6b). Evidently, the active energy at large $Pe$s could easily overcome the barrier of adsorption energy. The rapid detachment of ABPO leads to a short desorption time, which is analogue to a bounce of a compressed spring on a floor (without energy barrier). In this regime, the ABPO swells rapidly and $R_g^2(t)$ s including $R_{g\parallel}^2(t)$ s and $R_{g\perp}^2(t)$ s come up to a large value in a short time (as shown in Fig.S6). Besides, the desorption time is close to or smaller than $1.0\times10^2\tau$, where the ABPO displays a super-diffusion behavior rather than a normal diffusion (see Fig.S4(a-b)). At very small timescale, the velocity of center-of-mass is proportional to $\frac{F}{\gamma_T\sqrt{N}}$ for the continuous model [24-26]. Thus, the ABPO almost flies from the surface with a velocity proportional to active force, i.e., *Pe*.

For smaller *Pe*s ($\leq 32$), the desorption time also deviates from the exponential dependence. In this regime, the activation energy is week. The diffusion coefficient $D_\perp$ is smaller than $D_\parallel$ (Fig.3b and Fig.S4c). Due to the parameter close to the critical point, we did not find the scaling relationship $\tau_{des} \sim D_{eff}^{-1}$ predicted by Douglas *et al.*[67] for the passive polymer system. We find the desorption time in the regime displays a power-law scaling with $D_\perp$, $\tau_{des} \sim D_\perp^\beta$, here $\beta$ is scaling exponent close to $-1.28$ (see the inset of Fig.6b). The origin of the power-law relation and the scaling exponent is at present not clear, which motivates the development of new theories to understand the regime.

### 3.2 *Pe*-$\varepsilon_s$ phase diagram

To further understand the mutual competition of active force and attractive strength, we give the *Pe*-$\varepsilon_s$ phase diagram in Fig.7. At large $\varepsilon_s$ and low *Pe*, where the adsorption force is dominant, ABPO is in an adsorption state (see the phase diagram in Fig.7). In contrast, a desorption state exists at small $\varepsilon_s$ and high *Pe*, where the desorption force resulting from activity of ABPO becomes dominant. The two states are distinguished by $\tau_{des}$, obtained by fitting $\lg G(t) \sim t$. For an adsorption state, $\tau_{des}$ is too larger to be fitted. For a desorption state, $\tau_{des}$ is less than our simulation time ($4.0\times10^4\tau$). Additionally, an intermediate regime is defined when $\tau_{des}$ could be fitted but larger than our simulation time. With the increase of $\varepsilon_s$, the critical active force for polymer desorption also increases.

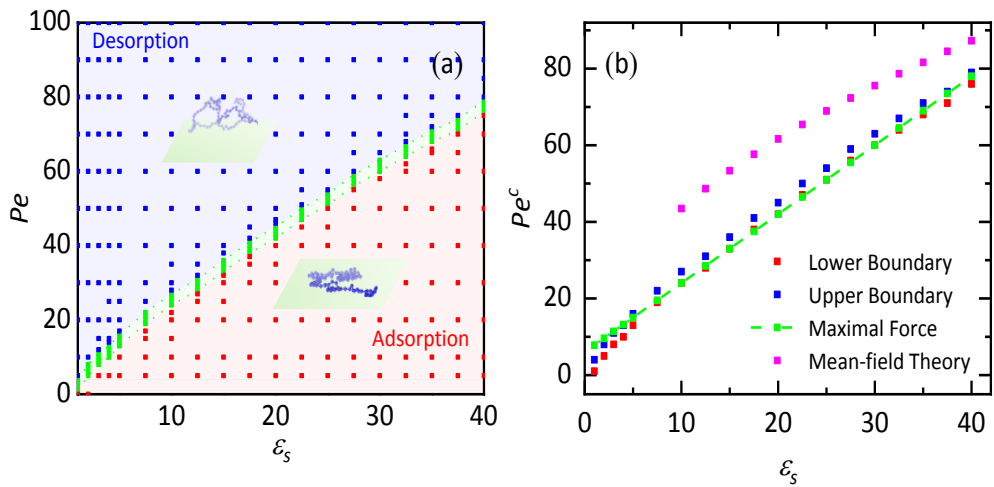

Fig.7. (a) $Pe$-$\varepsilon_s$ phase diagram for systems at $N=100$. (b) The critical Péclet number $Pe^c$ (critical phase boundary) as a function of $\varepsilon_s$ at $N=100$. The background colors are for eye guiding. The green color denotes the intermediate regime. Typical snapshots for adsorption ($Pe=15$, $\varepsilon_s=10$) and desorption ($Pe=35$, $\varepsilon_s=10$) states are displayed.

The phase boundary for $\varepsilon_s \geq 10$ can be fitted by the equation $Pe^c = 1.74\varepsilon_s + 7.20$ for lower boundary and $Pe^c = 1.74\varepsilon_s + 10.20$ for upper boundary (Fig.7b), which nearly linearly increases with attractive strength. In the strong attractive regime, the desorption of each monomer could be considered as a jump process. If one expects the jump process occurs frequently, the active force should be close to the maximal force, $-1.80\varepsilon_s/\sigma$, exerted by the attractive potential. The persistence time of each monomer is $(2D_R)^{-1}$. Thus, the critical Péclet number $Pe^c$ could be estimated by using the monomer locomotion over $\sim 1.0\sigma$ with a constant velocity equal to $(k_B T Pe^c/\sigma - 1.80\varepsilon_s/\sigma)/\gamma_T$ within the persistence time. Then, we get $Pe^c = 1.80\varepsilon_s + 6.00$. Additionally, we also use the mean-field theory (See SI) to roughly estimate the critical $Pe^c$, which is larger than the "maximal force" method (Fig.7b), which suggests that the critical desorption of the polymer is determined by the jump process of single monomer in the limit of strong desorption.

### 3.3 Effect of $D_R$

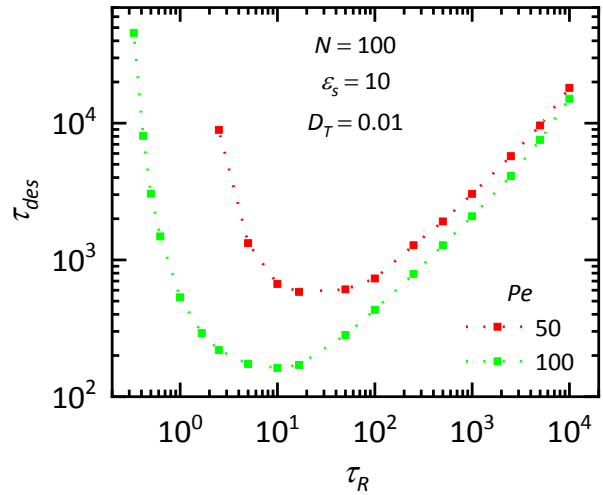

Fig. 8. $\tau_{des}$ as a function of $\tau_R = 1/2D_R$ for $Pe = 50, 100$ at $N=100$, $\varepsilon_s = 10$, $D_T = 0.01$.

To investigate the effect of rotation diffusion coefficient, we decouple the relation between $D_R$ and $D_T$, then carry out a series of simulations for $Pe = 50$ and $Pe = 100$, at $N=100$, $\varepsilon_s = 10$, $D_T = 0.01$. $\tau_{des}$ as a function of rotational relaxation time $\tau_R = 1/2D_R$ was plotted in Fig.8. It can be found that $\tau_{des}$ decreases first and then increases with $\tau_R$. The remarkable non-monotonic behavior demonstrates that, for a given $Pe$, there exists an optimal $D_R$ for the chain desorption. The phenomenon is similar to the escape of an Active Brownian Particle across a potential barrier.[71] For small $\tau_R$ s, the monomers rotate quickly. Not only the whole chain behaves like a passive polymer in a thermal bath with an effective temperature, each monomer also shows like a hot passive particle with effective translational diffusion coefficient $\propto Pe^2\tau_R$. That's why $\tau_{des}$ decreases as $\tau_R$ grows at first. For large $\tau_R$ s, each monomer hardly rotates in the time scale of $\tau_R$, thus there needs a long time waited for the occurrence of monomers' jumping, which will elongate the desorption time.

**3.4 Effect of chain length**

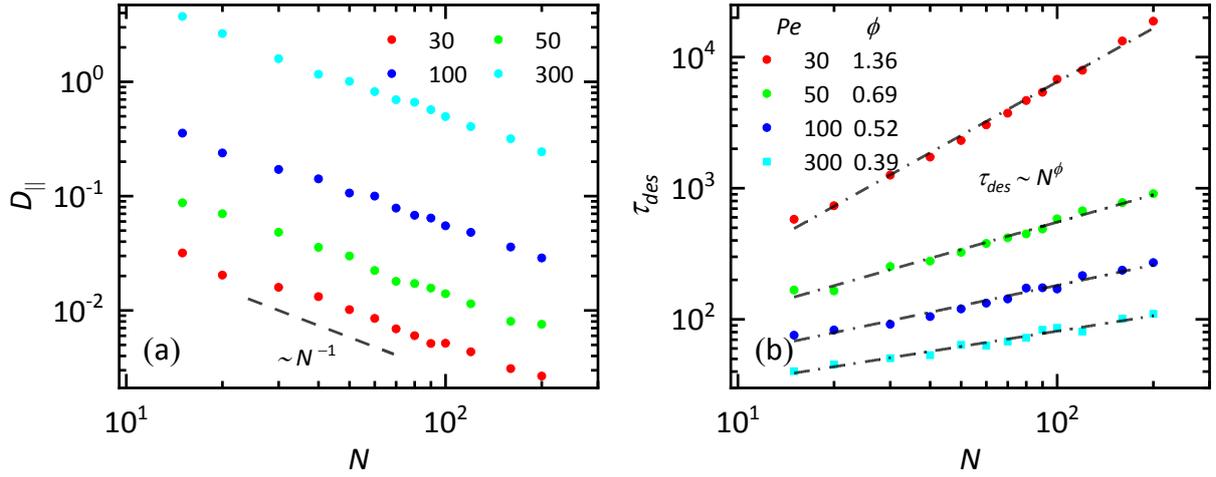

Fig.9. (a) The long-time diffusion coefficient $D_\parallel$ as a function of chain length $N$ for various $Pe$s with $\varepsilon_s = 10$. The dashed line is eye-guided power-law scaling relation $D_\parallel \sim N^{-1}$. (b) The desorption time $\tau_{des}$ as a function of chain length $N$ for various $Pe$s with $\varepsilon_s = 10$. The dashed lines are fitted for power-law scaling relation $\tau_{des} \sim N^\phi$.

The desorption dynamics is investigated with respect to the chain length $N$ ranging from 15 to 200 for systems at $Pe = 30, 50, 100, 300$. The long-time diffusion coefficients $D_\parallel$ versus $N$ are given in Fig.9a. It can be found that there exists a scaling relation $D_\parallel \propto N^{-1}$, for different $Pe$s, being consistent with the predication of active Rouse dynamics due to the $D \sim \frac{k_B T_{eff}}{\gamma_T N}$.[73] However, it should be noticed that the short-time dynamics of ABPOs also displays a super-diffusion behavior ($MSD_\parallel(t)$ for various $Pe$s with different chain length shown in Fig.S7). Further, we find $\tau_{des}$ increases in a power-law relation with $N$, $\tau_{des} \sim N^\phi$. $\phi$ are 1.36, 0.69, 0.52 and 0.39 for corresponding $Pe$s, respectively (Fig.9b). High $Pe$ results in a small $\phi$. This is attributed to the mutual competition between active force and attractive force. Imaginably, the exponent might be close to zero, when the active force is large enough. This infers that the desorption time is less dependent on the chain length, as a result of active force easily overcoming the attractive energy from the surface. Furthertly, the scaling exponent $\phi$ is smaller than that given by Paturej's et al.[48]. This reveals that a larger $Pe$ coupling a swelling conformation can accelerate the desorption process. In addition, we also find that the adsorption and desorption states are nearly not dependent the chain length. The $Pe$ parameters are same for adsorbed and desorbed regions regardless of chain length (see Fig.S8). The results might originate from the independent active forces exerted on each monomer and the "maximal force" should be overcome. Also, the exponential dependence of desorption time on the inverse of *effective temperature* is obtained at a certain range of *Pe*s, except that $\alpha$ s depend on the chain length (smaller for the shorter ABPO, See Fig.S9).

## 4. Conclusion and Discussion

We explored the conformational and diffusive behaviors of a flexible active Brownian polymer (ABPO) on a flat attractive surface by computer simulations. The phase diagram with respect to activity and adsorption strength was obtained. At large adsorption strength and low activity, an adsorption state with ABPO spreading and diffusing on the surface exits. While, small adsorption strength and high activity give rise to a desorption state. At a certain adsorption strength, the adsorption-desorption transition occurs with the increase of activity. The diffusion behavior of ABPO parallel to the surface is in good agreement with that predicated by 3D active Rouse model. Hence, an effective temperature, $T_{eff}$ can be defined due to the existence of non-thermal active noise.

Furthermore, we paid attention to its desorption kinetics by extracting the desorption time by linear fitting $\lg G(t) \sim t$. Our results reveals that the desorption time $\tau_{des}$ and $T_{eff}$ follows an exponential decay relation, as $\tau_{des} \sim \exp(\alpha/k_B T_{eff})$ at $32 \leq Pe \leq 120$, similar to the mechanism of thermal activation. Interestingly, we also found the deviation of desorption time from the exponential decay at *Pe*s larger and smaller *Pe*s, which might be explained by activity-induced bounce and diffusion-like process. Moreover, we investigated the effect of rotational diffusion coefficient and polymer length. We observed a non-monotonic dependence of desorption time on the $D_R$ and perceived there exists an optimal $D_R$ for the chain desorption, similar to the escape of an Active Brownian Particle across a potential barrier. Additionally, we found the desorption time increases in a power-law relation with chain length, i. e., $\tau_{des} \propto N^\phi$. The mutual competition between active force and attractive force contributes to the decrease of $\phi$ with the increase of active strength. Our results highlight that the active force can be used to regulate the desorption behavior of polymers.

Experimentally, it is difficult to realize the desorption phenomena via using the synthetic polymers and the strength of active coupling of bio-macromolecules mediated by binding enzymes is also hard to be tuned.[50] Recently, the concept of colloidal molecules, consisting of several species of colloidal particles tightly bound together in mesoscopic scale, was proposed [74, 75] Active colloidal polymers are composed of "active" or "self-propelled" colloids.[76] The desorption of "active colloidal polymer" from a solid surface can be realized via controlling the activity of colloids, which, in principle, could be used to validate the results we observed. Although the colloidal polymers are at the different spatial scale with traditional polymers, the physical picture is similar because of the similar Langevin equation they obey.

It also should be noted that this is the first step to understand the activity-induced desorption of polymers. There is only one flexible polymer in our simulation system, which corresponds to a very dilute solution. The concentration of ABPO and its stiffness, which affects the configurational entropy of ABPO, will be considered in further work. Another interesting question is exploring the effect of translational friction coefficient, i.e., the effect of inertia. In our cases, the characteristic timescale for inertial effect is about $m/\gamma_T = 10^{-2}\tau$, which is very shorter than the desorption time. Thus, the inertia effect could be ignored. When the inertial characteristic timescale is close to and larger than desorption time, the inertia effect should be concerned.[12, 14, 55-59]

**Conflicts of interest**

There are no conflicts to declare.

**Supporting Information**

The Supporting Information is available free of charge at xxxxxxx.

The re-adsorption behavior, a Schematic diagram for calculating the time-dependent and time-independent quantities, the mean-square radius of gyration, mean-square displacement and diffusion coefficient, the Mean-field theory, and Figures for N effect.

**Acknowledgements**

This work was supported by the National Natural Science Foundation of China (NSFC) of Grant No. 21674078.